\documentclass[a4paper]{article}
\usepackage{Odyssey2024}
\ninept

\usepackage{dblfloatfix}    
\usepackage{listings}
\usepackage{hyperref}
\usepackage{adjustbox}
\usepackage{amsmath}
\usepackage{amsthm}
\usepackage{amssymb}
\usepackage{mathrsfs}
\usepackage{bigdelim}

\usepackage{wrapfig}
\usepackage[labelformat=simple]{subcaption}

\usepackage[font=small,labelfont=bf]{caption}
\usepackage{bbm}
\usepackage{pifont}
\usepackage{enumitem}
\usepackage{multirow}
\usepackage{adjustbox}
\usepackage{booktabs}
\usepackage[hang,flushmargin]{footmisc} 

\usepackage[T1]{fontenc}

\usepackage{pifont}
\newcommand{\cmark}{\ding{51}}%
\newcommand{\xmark}{\ding{55}}%

\usepackage{xcolor}
\usepackage[linesnumbered,ruled,vlined]{algorithm2e}
\allowdisplaybreaks

\usepackage[labelformat=simple]{subcaption}

\usepackage[font=small,labelfont=bf]{caption}

\usepackage[T1]{fontenc}

\SetCommentSty{mycommfont}

\SetKwInput{KwInput}{Input}                
\SetKwInput{KwOutput}{Output}              
\SetKwInput{KwInit}{Init.}              

\renewenvironment{thebibliography}[1]{
  \begin{oldthebibliography}{#1}
    \setlength{\itemsep}{0em}
    \setlength{\parskip}{0em}
}
{
  \end{oldthebibliography}
}

\usepackage{tikz}

\title{On Speaker Attribution with SURT}
\name{\begin{tabular}{c}
Desh Raj$^1$\thanks{This work was partially funded by the National Science Foundation CCRI
program via Grant No. 2120435.}, Matthew Wiesner$^2$, Matthew Maciejewski$^2$, Leibny Paola Garc\'{\i}a-Perera$^{1,2}$,\\Daniel Povey$^3$, Sanjeev Khudanpur$^{1,2}$
\end{tabular}}
\address{
  $^1$CLSP \& $^2$HLTCOE, Johns Hopkins University, Baltimore, USA; $^3$Xiaomi Corp., Beijing, China \\
{\small \tt draj@cs.jhu.edu, \{wiesner, mmaciej2, leibny\}@jhu.edu, dpovey@gmail.com, khudanpur@jhu.edu}
}

\begin{document}

\setlength{\abovedisplayskip}{2pt}
\setlength{\belowdisplayskip}{2pt}

\maketitle
\begin{abstract}
The Streaming Unmixing and Recognition Transducer (SURT) has recently become a popular framework for continuous, streaming, multi-talker speech recognition (ASR).
With advances in architecture, objectives, and mixture simulation methods, it was demonstrated that SURT can be an efficient streaming method for \textit{speaker-agnostic} transcription of real meetings.
In this work, we push this framework further by proposing methods to perform \textit{speaker-attributed} transcription with SURT, for both short mixtures and long recordings.
We achieve this by adding an auxiliary speaker branch to SURT, and synchronizing its label prediction with ASR token prediction through HAT-style blank factorization.
In order to ensure consistency in relative speaker labels across different utterance groups in a recording, we propose ``speaker prefixing'' --- appending each chunk with high-confidence frames of speakers identified in previous chunks, to establish the relative order.
We perform extensive ablation experiments on synthetic LibriSpeech mixtures to validate our design choices, and demonstrate the efficacy of our final model on the AMI corpus.
\end{abstract}

\section{Introduction}
\label{sec:intro}

Speaker-attributed multi-talker speech recognition (ASR), or ``who spoke what'', is the task of transcribing all the speech in a multi-talker conversation along with relative speaker attribution.
This task has several applications such as meeting transcription and summarization, collaborative learning, and dinner-party conversations~\cite{Barker2015TheT, Kinoshita2013TheRC, Watanabe2020CHiME6CT}.
Due to the presence of overlapping speech, turn-taking, and far-field audio, it often requires special modeling techniques~\cite{Carletta2005TheAM, Shriberg2001ObservationsOO, Yoshioka2019MeetingTU}. 
Researchers have worked on speaker-attributed transcription from modular (i.e., pipeline-based) and end-to-end perspectives.
In the former, it is decomposed into speaker diarization and ASR sub-tasks and addressed independently, leveraging advances in each of these fields~\cite{Raj2022GPUacceleratedGS,Raj2020IntegrationOS,Kanda2019SimultaneousSR}.
However, this approach may be sub-optimal since the components are independently optimized leading to error propagation, and may also require greater engineering efforts for maintenance~\cite{Wu2021InvestigationOP}. 

Due to these limitations with modular systems, researchers have proposed jointly optimized models that combine diarization and ASR to directly solve for speaker-attributed transcription. 
The most popular of these is speaker-attributed ASR (SA-ASR) based on attention-based encoder-decoders (AEDs)~\cite{Chorowski2015AttentionBasedMF}.
It uses serialized output training (SOT) to handle overlapped speech and registered speaker profiles (called a speaker inventory) to handle speaker attribution~\cite{Kanda2020SerializedOT,Kanda2020JointSC}.
Several modifications to this model have leveraged transformer-based encoders~\cite{Kanda2021EndtoEndSA} and large-scale pre-training~\cite{Kanda2021ACS}, and have proposed methods for inference on long recordings~\cite{Chang2021HypothesisSF} without the dependence on speaker inventory~\cite{Kanda2020InvestigationOE}.
There have been further investigations on methods for speaker attribution within SA-ASR, and its extension to multi-channel and contextualized ASR~\cite{Yu2022ACS,Shi2022ACS,Shi2023CASAASRCS}.
By modifying SOT to be performed at the token-level (known as t-SOT), Kanda et al.~\cite{Kanda2022StreamingSA} performed streaming transcription of overlapping speech, which was not feasible with utterance-level serialization.
Enforcing monotonicity in this manner also allows these models to be built upon neural transducers~\cite{Graves2012SequenceTW} instead of AEDs.
Nevertheless, t-SOT requires complicated interleaving/deserialization of tokens based on timestamps to accommodate overlapping speech on a single output channel, and the use of ``channel change'' tokens may impact ASR training adversely.
%

\begin{figure}
    \centering
    \includegraphics[trim={0.5cm 0 0 0},clip,width=\linewidth]{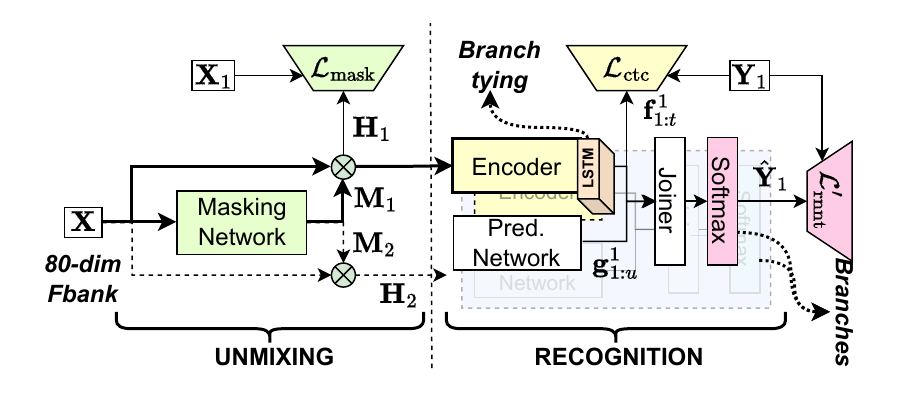}
    \vspace{-2em}
    \caption{An overview of SURT 2.0, as described in \cite{Raj2023Surt20}. It consists of a masking network and a transducer-based ASR.}
    \label{fig:surt}
\end{figure}

An alternative approach for continuous, streaming, multi-talker ASR involves transcribing overlapping utterances on parallel output channels by unmixing them inside the model.
This two-branch strategy is exemplified by models such as Streaming Unmixing and Recognition Transducer (SURT)~\cite{Lu2020StreamingEM} and multi-turn RNN-T (MT-RNNT)~\cite{Sklyar2021MultiTurnRF}, but we will refer to them as SURT in this paper without loss of generality.
SURT has been extended to handle long-form multi-turn recordings~\cite{Raj2021ContinuousSM,Sklyar2021MultiTurnRF}, and to jointly perform endpointing and segmentation~\cite{Lu2022EndpointDF,Sklyar2022SeparatorTransducerSegmenterSR}.
Lu et al.~\cite{Lu2021StreamingMS} also proposed joint speaker identification with SURT, but their model relied on a speaker inventory and was only used for single-turn synthetic mixtures.
As shown in Fig.~\ref{fig:surt}, the SURT model consists of an ``unmixing'' component that separates the mixed audio into non-overlapping streams, and a ``recognition'' component that transcribes each of these streams.
Since there is no explicit emission of speaker labels in this modeling scheme, SURT has thus far been limited to \textit{speaker-agnostic} transcription.
In this paper, our objective is to extend the SURT model for \textit{speaker-attributed} transcription of an arbitrary number of speakers without any speaker inventory.

We achieve this by adding an auxiliary speaker transducer to the recognition module of SURT.
We constrain this branch to emit a speaker label corresponding to each ASR token predicted by using HAT-style blank factorization of the output logits.
We also propose a novel ``speaker prefixing'' method to ensure that the speaker labels are consistent across different utterance groups in the recording.
We validate our methods through ablation experiments on LibriSpeech mixtures, and finally demonstrate streaming speaker-attributed transcription on real meetings from the AMI corpus.
We will publish code through the open-source \texttt{icefall} toolkit\footnote{\url{https://github.com/k2-fsa/icefall}}.

\section{Preliminary: SURT}

\subsection{Speech recognition with neural transducers}
\label{sec:asr}

In single-talker ASR, audio features for a segmented utterance $\mathbf{X} \in \mathbb{R}^{T\times F}$ ($T$ and $F$ denote the number of time frames and the input feature dimension, respectively) are provided as input, and the system predicts the transcript $\mathbf{y} = (y_1,\ldots,y_U)$, where $y_u \in \mathcal{V}$ are output units such as graphemes or word-pieces, and $U$ is the length of the label sequence. 
For discriminative training, we achieve this by minimizing the negative conditional log-likelihood, $\mathcal{L} = -\log P(\mathbf{y}|\mathbf{X})$. 
%
%
Since the alignment between $\mathbf{X}$ and $\mathbf{y}$ is not known, transducers compute $\mathcal{L}$ by marginalizing over the set of all alignments $\mathbf{a} \in \bar{\mathcal{V}}^{T+U}$, where $\bar{\mathcal{V}} = \mathcal{V}\cup \{\phi\}$ and $\phi$ is called the blank label.
Formally,
\begin{equation}
P(\mathbf{y}|\mathbf{X}) = \sum_{\mathbf{a}\in \mathcal{B}^{-1}(\mathbf{y})} P(\mathbf{a}|\mathbf{X}),
\label{eq:rnnt}
\end{equation}
where $\mathcal{B}$ is a deterministic mapping from an alignment $\mathbf{a}$ to an output sequence $\mathbf{y}$. 
%
%
Transducers parameterize $P(\mathbf{a}|\mathbf{X})$ with an encoder, a prediction network, and a joiner (see ``recognition'' component in Fig.~\ref{fig:surt}).
The encoder maps $\mathbf{X}$ into hidden representations $\mathbf{f}_1^T$, while the prediction network maps $\mathbf{y}$ into $\mathbf{g}_1^U$.
The joiner combines the outputs from the encoder and the prediction network to compute logits $\mathbf{z}_{t,u}$ which are fed to a softmax function to produce a posterior distribution over $\bar{\mathcal{V}}$. 
Under the assumption of a streaming encoder, we can expand \eqref{eq:rnnt} as
\begin{align}
    P(\mathbf{y}|\mathbf{X}) &= \sum_{\mathbf{a}\in \mathcal{B}^{-1}(\mathbf{y})} \prod_{t=1}^{T+U} P(\mathbf{a}_t|\mathbf{f}_1^t,\mathbf{g}_1^{u(t)-1}) \\
    &= \sum_{\mathbf{a}\in \mathcal{B}^{-1}(\mathbf{y})} \prod_{t=1}^{T+U} \mathrm{Softmax}(\mathbf{z}_{t,u(t)}), \label{eq:rnnt_softmax}
\end{align}
where $u(t)\in\{1,\ldots,U\}$ denotes the index in the label sequence at time $t$.
The negative log of this expression is known as the RNN-T or transducer loss.
In practice, to make training more memory-efficient, we often approximate the full sum, for example using the pruned transducer loss~\cite{Kuang2022PrunedRF}.
We will denote this loss as $\mathcal{L}_{\text{rnnt}}$ for the remainder of this paper.

\subsection{Multi-talker ASR with SURT}
\label{sec:mt-asr}

In multi-talker ASR, the input $\mathbf{X}\in\mathbb{R}^{T\times F}$ is an unsegmented mixture containing $N$ utterances from $K$ speakers, i.e., $\mathbf{X} = \sum_{n=1}^N \mathbf{x}_n$, where $\mathbf{x}_n$ is the $n$-th utterance ordered by start time, shifted and zero-padded to the length of $\mathbf{X}$. 
The desired output is $\mathbf{Y} = \{\mathbf{y}_n: 1\leq n \leq N\}$, where $\mathbf{y}_n$ is the reference corresponding to $\mathbf{x}_n$. 
Assuming at most two-speaker overlap, the \textit{heuristic error assignment training} (HEAT) paradigm~\cite{Lu2020StreamingEM} is used to create channel-wise references $\mathbf{Y}_1$ and $\mathbf{Y}_2$ by assigning $\mathbf{y}_n$'s to the first available channel, in order of start time. 
SURT estimates $\hat{\mathbf{Y}} = [\hat{\mathbf{Y}}_1,\hat{\mathbf{Y}}_2] = f_{\text{surt}}(\mathbf{X})$ as follows. 
First, an unmixing module computes $\mathbf{H}_1$ and $\mathbf{H}_2$ as
\begin{align}
\begin{split}
\label{eq:surt}
& \mathbf{H}_1 = \mathbf{M}_1 \ast \mathbf{X}, \quad \mathbf{H}_2 = \mathbf{M}_2 \ast \mathbf{X},~~\text{where} \\
& [\mathbf{M}_1,\mathbf{M}_2]^T = \mathrm{MaskNet}(\mathbf{X}),
\end{split}
\end{align}
$\mathbf{M}_c\in \mathbb{R}^{T\times F}$ is a soft mask per channel and $\ast$ is Hadamard product. 
$\mathbf{H}_1$ and $\mathbf{H}_2$ are fed into a transducer-based ASR, producing logits $\mathbf{Z}_1$ and $\mathbf{Z}_2$. 
Finally,
\begin{align}
\begin{split}
\label{eq:heat}
&\mathcal{L}_{\text{heat}} = \mathcal{L}(\mathbf{X}, \mathbf{Y}_1, \mathbf{Z}_1) + \mathcal{L}(\mathbf{X}, \mathbf{Y}_2, \mathbf{Z}_2),~~\text{where}\\
&\mathcal{L} = \mathcal{L}_{\text{rnnt}} + \lambda_{\text{ctc}}\mathcal{L}_{\text{ctc}} + \lambda_{\text{mask}}\mathcal{L}_{\text{mask}},
\end{split}
\end{align}
where $\mathcal{L}_{\text{ctc}}$ and $\mathcal{L}_{\text{mask}}$ denote auxiliary CTC loss on the encoder~\cite{Graves2006ConnectionistTC} and mean-squared error loss on the masking network, respectively, and $\lambda$'s are hyperparameters. 
In the original formulation, SURT only performs \textit{speaker-agnostic} transcription, and is evaluated using ORC-WER (cf. Section~\ref{sec:evaluation})~\cite{Sklyar2021MultiTurnRF}.

\section{Methodology}


For speaker-attributed ASR, the desired output is $\mathbf{Y}=\{(\mathbf{y}_n,s_n):1\leq n \leq N, s_n \in [1,K]\}$, where $K$ is the number of speakers in the mixture.
SURT estimates $\mathbf{y}_n$ by mapping the utterances to two channels $\hat{\mathbf{Y}}_1$ and $\hat{\mathbf{Y}}_2$, as described in Section~\ref{sec:mt-asr}.
A popular method for speaker attribution in multi-talker settings is to predict speaker change tokens that segment the output into speaker-specific regions, followed by speaker label assignment to each segment.
However, this kind of training is prone to over-estimate the speaker change tokens, and may also adversely affect the ASR performance.
Instead, we want to perform speaker attribution without affecting the output of the ASR branch, for example by predicting a speaker label for each ASR token emitted.

In order to perform such a streaming speaker attribution jointly with the transcription, the following questions arise: 
(i) How do we deal with overlapping speech? 
(ii) How do we synchronize speaker label prediction with ASR token prediction? 
(iii) How to reconcile relative speaker labels across utterance groups in a long recording?
We will answer each of these questions in the following subsections.

\subsection{Auxiliary speaker transducer}

\begin{figure}[t]
    \centering
    \includegraphics[width=0.9\linewidth]{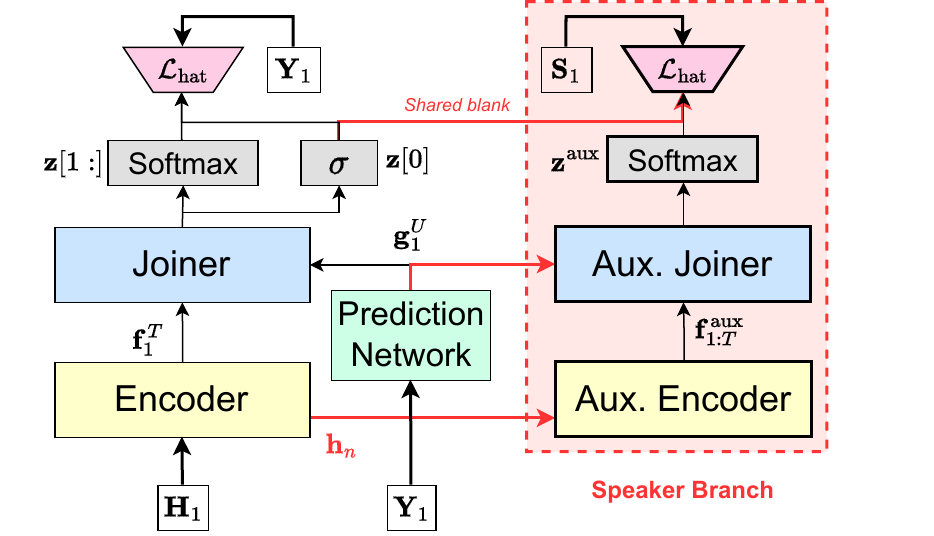}
    \vspace{-1em}
    \caption{Auxiliary speaker transducer (red box) with shared blank label. The auxiliary encoder takes as input a hidden layer representation $\mathbf{h}_n$ from the main encoder, and generates $\mathbf{f}_{1:T}^{\mathrm{aux}}$. The blank logit $\mathbf{z}[0]$ from the main joiner is shared with the speaker branch to compute the HAT loss.}
    \label{fig:aux_branch}
\end{figure}

\begin{figure*}[tbp]
\centering
    \captionsetup[subfigure]{labelformat=empty}
    \begin{subfigure}{0.3\linewidth}
    \centering
    \includegraphics[width=\linewidth]{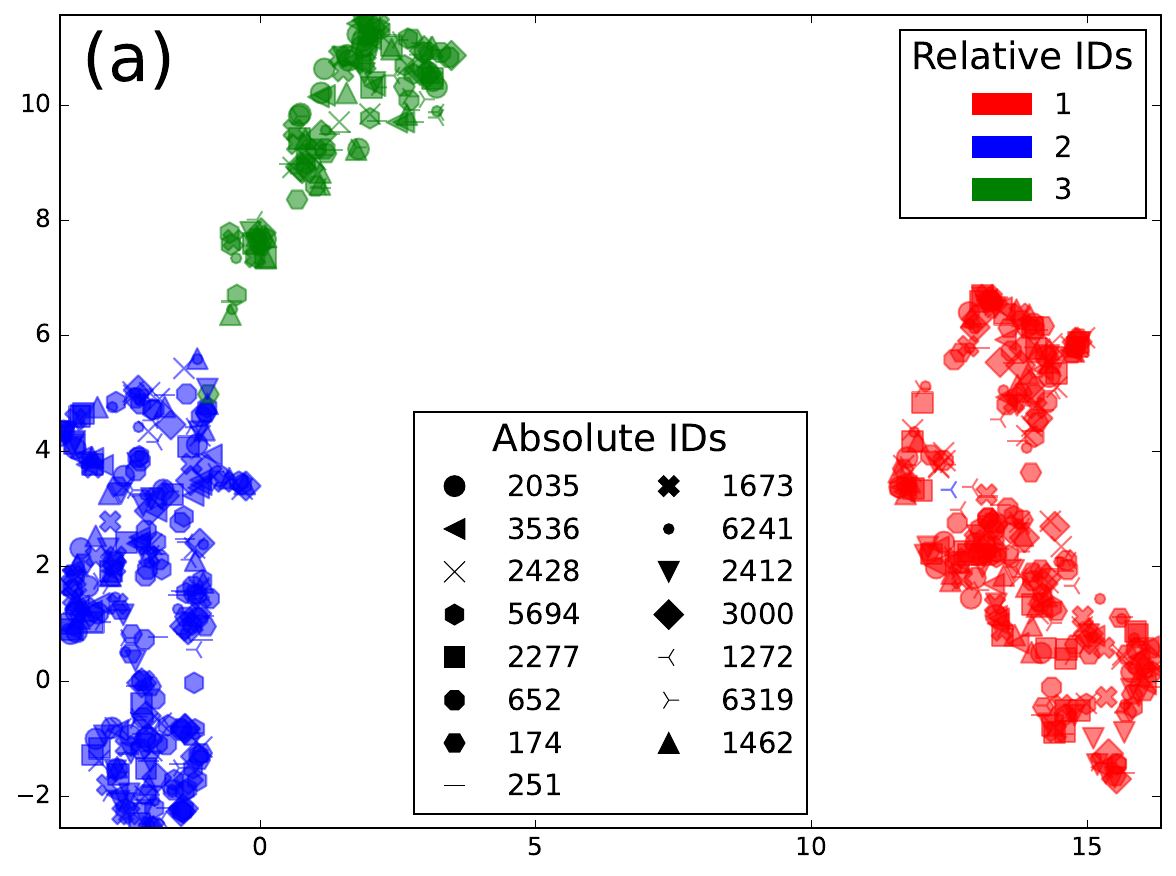}
    \caption{\label{fig:umap}}
    \end{subfigure}
    \begin{subfigure}{0.3\linewidth}
    \centering
    \includegraphics[width=\linewidth]{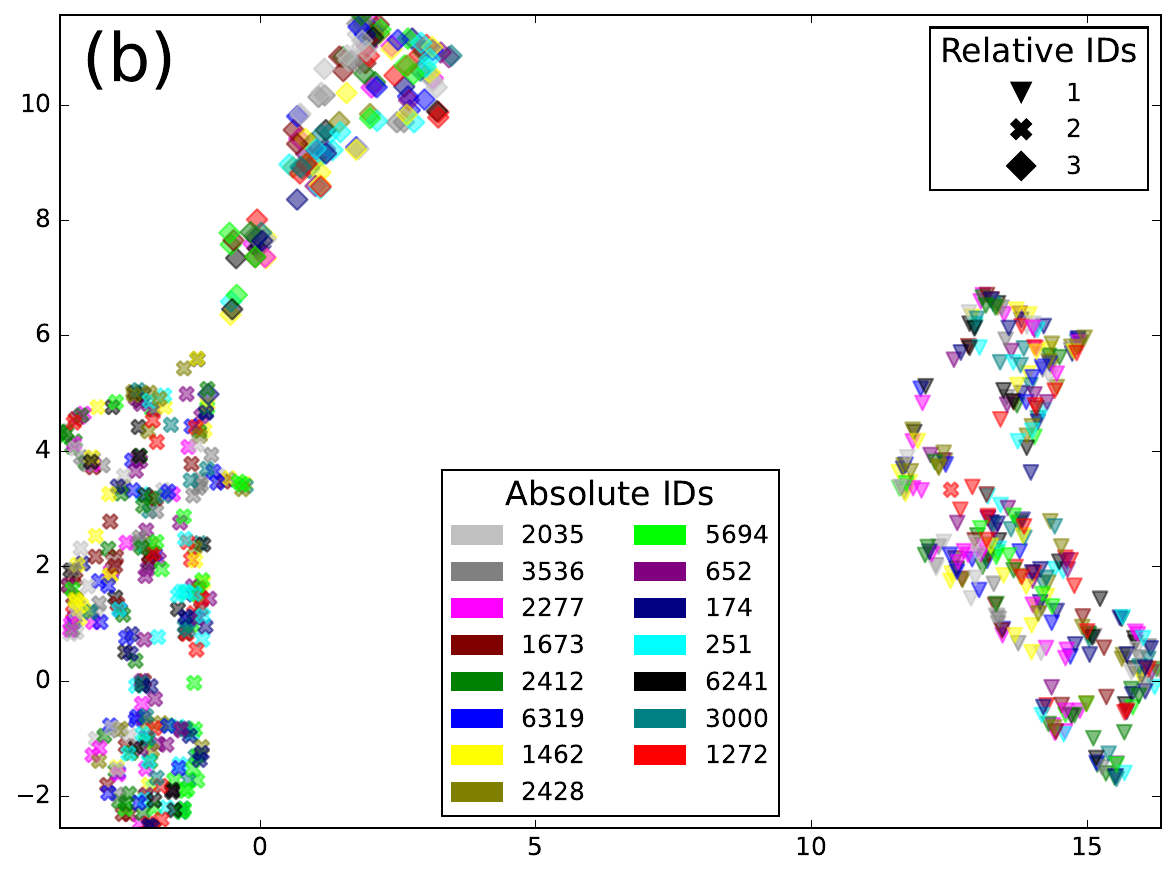}
    \caption{\label{fig:umap_inv}}
    \end{subfigure}
    \begin{subfigure}{0.3\linewidth}
    \centering
    \includegraphics[width=\linewidth]{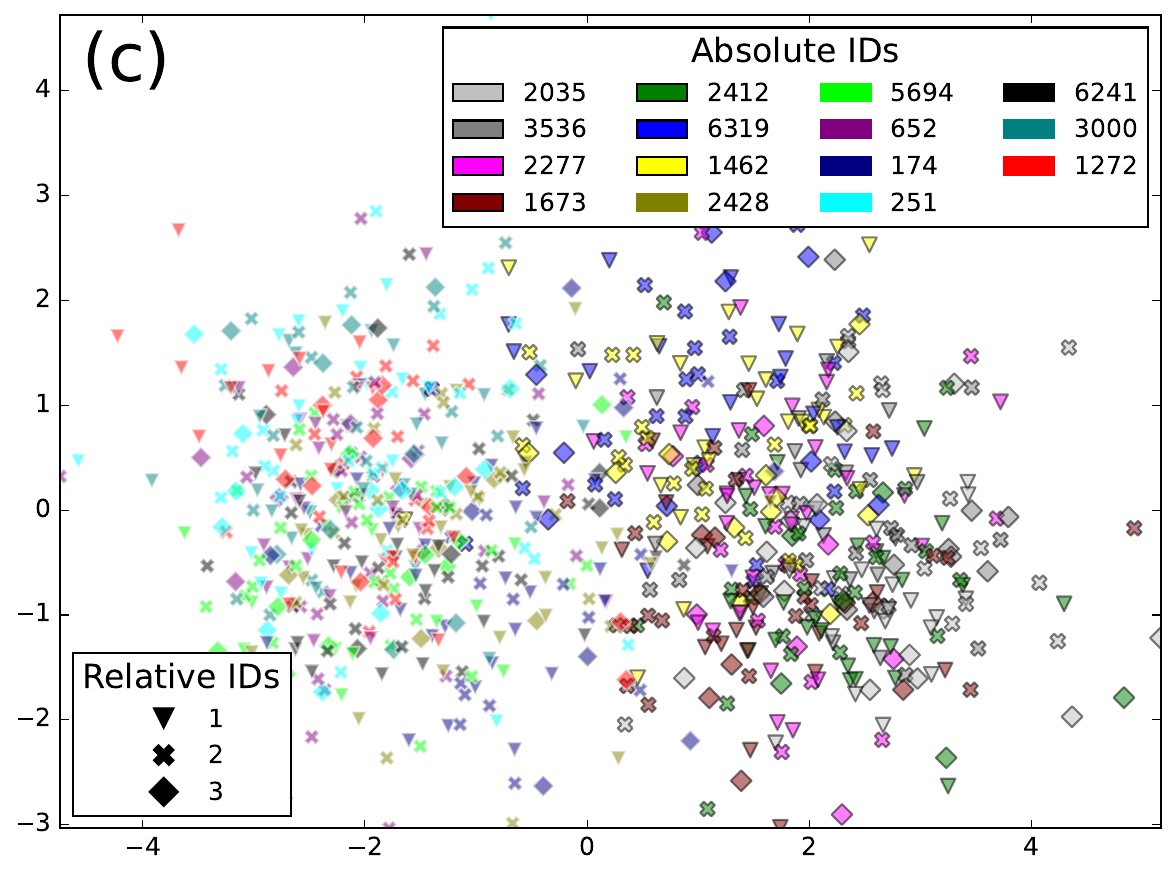}
    \caption{\label{fig:lda_inv}}
    \end{subfigure}
\vspace{-2em}
\caption{Projections of auxiliary encoder representations for a subset of LSMix \texttt{dev}. Each point denotes the representation of one speaker in a mixture, averaged over the frames on which the model emits a non-blank label. (a) and (b) denote UMAP projection, and (c) shows LDA projection using absolute speaker classes.}
\label{fig:spk_clusters}
\vspace{-2em}
\end{figure*}

We map speaker labels $s_n$ to two channels according to the HEAT strategy, obtaining $\hat{\mathbf{S}}_1$ and $\hat{\mathbf{S}}_2$.
During training, we repeat $s_n$ as many times as there are tokens in $\mathbf{y}_n$, i.e., we want to predict a speaker label for each ASR token.
Thereafter, we use the non-overlapping streams $\mathbf{H}_c$ to estimate $\hat{\mathbf{S}}_c$ in the same two-branch approach as the ASR transducer.
For this, we add an auxiliary speaker transducer to each of the two branches in the recognition module, as shown in Fig.~\ref{fig:aux_branch}.
Intermediate representations $\mathbf{h}_n$ from the $n^{\text{th}}$ layer of the main encoder are fed into an auxiliary encoder, producing $\mathbf{f}_{1:T}^{\mathrm{aux}}$.
An auxiliary joiner combines $\mathbf{f}_{1:T}^{\mathrm{aux}}$ with $\mathbf{g}_{1}^U$ to produce auxiliary logits $\mathbf{z}_{t,u}^{\mathrm{aux}}$, which are used to obtain a distribution over the speaker labels and the blank label.
Combing the auxiliary encoder representation with representations from the ASR prediction network allows the speaker branch to leverage lexical content for predicting speaker labels.
Such a use of lexical information has been shown to be beneficial for speaker diarization using clustering-based~\cite{Flemotomos2019LanguageAS, Park2019SpeakerDW} or end-to-end neural approaches~\cite{Khare2022ASRAwareEN}.

\subsection{Synchronizing speaker labels with ASR tokens}

Since transducers perform frame-synchronous decoding with the blank label, the above formulation has several issues.
First, we cannot ensure that the number of ASR tokens $\lvert \hat{\mathbf{Y}}_c\rvert$ predicted on branch $c$ is equal to the the number of speaker labels $\lvert \hat{\mathbf{S}}_c \rvert$.
Even if we can ensure this, assigning speaker labels to ASR tokens can be hard, as evident from the following example containing two speakers saying the words ``hello'' and ``hi'':
\begin{figure}[h]
    \centering
    \includegraphics[width=0.9\linewidth]{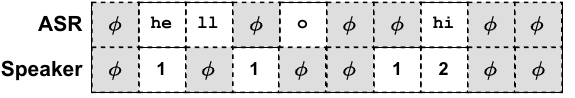}
\end{figure}

Even though we predicted the correct speaker labels, it is hard to assign them to the corresponding ASR tokens since they are not synchronized by frame.
To solve these problems, we need to ensure that SURT emits blank labels on the same frames for both the ASR and speaker branches.
We achieve this by factoring out the blank label separately in the style of the hybrid auto-regressive transducer (HAT) model~\cite{Variani2020HybridAT}, i.e., we replace the alignment posterior $P(\mathbf{a}_t\mid \mathbf{f}_1^t, \mathbf{g}_1^{u(t)-1})$ in \eqref{eq:rnnt_softmax} with
\begin{equation}
\label{eq:hat}
\begin{cases} 
b_{t,u}, ~~\text{if}~~ \mathbf{a}_t = \phi, \\
(1-b_{t,u})~\mathrm{Softmax(\mathbf{z}_{t,u}[1:])}, ~~\text{otherwise},
\end{cases}
\end{equation}
where $b_{t,u} = \sigma(\mathbf{z}_{t,u}[0])$, and $\sigma$ denotes the sigmoid function.
By setting $\mathbf{z}_{t,u}^{\mathrm{aux}}[0] = \mathbf{z}_{t,u}[0]$, i.e., by sharing the blank logit for the ASR and speaker outputs, we ensure that blank emission is synchronized between the two branches.
The speaker branch is trained with a similar HAT loss, i.e.,
\begin{equation}
\label{eq:spk_hat}
\mathcal{L}_{\mathrm{aux}} = \mathcal{L}_{\mathrm{hat}}(\mathbf{H}_1,\mathbf{Z}_1^{\mathrm{aux}}) + \mathcal{L}_{\mathrm{hat}}(\mathbf{H}_2,\mathbf{Z}_2^{\mathrm{aux}}).
\end{equation}
Such a synchronization strategy has also recently been proposed for performing word-level diarization using transducers~\cite{Huang2023TowardsWE}.
For both ASR and speaker branches, we use a pruned version of the HAT loss similar to pruned RNNT~\cite{Kuang2022PrunedRF}.

\subsection{Maintaining state across utterance groups}


A common approach for inference of long-form audio is by chunking in some way (e.g., at silences or fixed-length chunks), processing each chunk separately, and then combining the outputs.
For SURT, we assume that the recording has been chunked at silences to create utterance groups, which are sets of utterances connected by speaker overlaps.
For multi-talker ASR methods such as SA-ASR (c.f. Section~\ref{sec:intro}) which predict \textit{absolute} speaker identities using external speaker profiles, combining chunk-wise outputs is relatively straightforward since there is no issue of speaker label permutation.
However, the auxiliary speaker branch in SURT is trained to predict \textit{relative} speaker labels in FIFO order within that chunk, and these labels must be reconciled across all chunks within a recording in order to obtain the final speaker-attributed transcript.
%

\subsubsection{What does the auxiliary encoder encode?}

Speaker label reconciliation across different chunks for long-form diarization or speaker-attributed ASR is often done through clustering of speaker embeddings estimated from the chunks.
For example, the EEND-VC model for speaker diarization extends EEND for diarization of long-recordings by applying clustering over chunk-wise speaker vectors~\cite{Kinoshita2020IntegratingEN}.
This method delays the output prediction at least until the end of the chunk so that the re-clustering may be done.
To remedy this issue, SA-ASR based on t-SOT estimates speaker change based on cosine similarity between consecutive speaker vectors, and applies re-clustering of all vectors every time a speaker change is detected~\cite{Kanda2022StreamingSA}.
Nevertheless, solving label permutation through such clustering requires that the chunk-wise speaker vectors should represent absolute speaker identities.
This requirement may not be satisfied in the SURT model since the auxiliary speaker branch is trained to predict relative speaker labels in their order of appearance in the mixture.

To verify this, we applied SURT with the auxiliary speaker branch on synthetic mixtures of LibriSpeech utterances (described in Section~\ref{sec:data}) consisting of 2 or 3 speakers per mixture.
We collected the 256-dimensional encoder representations for the frames where the auxiliary branch predicts a speaker label, and averaged each speaker's embeddings over the mixture.
In Fig.~\ref{fig:spk_clusters}, we show UMAP and LDA projections of these embeddings for 15 different speakers in the LSMix \texttt{dev} set.
In Fig.~\ref{fig:umap}, the three colors denote the relative speaker label assigned to the speaker during SURT inference and the markers denote absolute speaker identities.
Fig.~\ref{fig:umap_inv} shows the same plot, but in this case colors denote absolute speaker identities and markers denote relative order within the chunk.
It is easy to see that the embeddings cluster by relative speaker labels instead of absolute speaker identities, validating our conjecture that the auxiliary encoder extracts relative speaker position in the chunk.
Even when LDA using absolute speaker labels is used for the low-dimensional projection (as shown in Fig.~\ref{fig:lda_inv}), we did not find clusters of absolute speaker labels.
Interestingly, the embeddings did retain information about the speaker's gender. 
In the figure, the points with and without a black border denote female and male speakers, respectively, and they appear well-separated into gender-based clusters.

\subsubsection{The speaker prefixing method}

Inspired by the use of a speaker tracing buffer in the EEND model for online diarization~\cite{Xue2020OnlineEN}, we propose a novel \textit{speaker prefixing} strategy to solve the problem of speaker label permutation across utterance groups.
The idea of speaker prefixing is to append high-confidence frames for speakers we have seen so far in the recording, before the chunk's input features, in the order of their predicted label.
Formally, let $\{\mathbf{X}_1,\ldots,\mathbf{X}_M\}$ be the input features corresponding to $M$ utterance groups in a recording, such that $\mathbf{X}_m \in \mathbb{R}^{T_m \times F}$.
For some chunk $m$, let $K_m \in [0,K]$ be the number of speakers seen so far in the recording.
We define some function $\mathscr{S}$ which selects frames of a given speaker in the previous chunks, i.e.,
\begin{equation}
    \mathscr{S}(\mathbf{X}_1,\ldots,\mathbf{X}_{m-1},k) = \mathbf{B}_k,
\end{equation}
where $k$ is one of the $K_m$ speakers and $\mathbf{B}_k \in \mathbb{R}^{\tau \times F}$, for some $\tau$ (which is a hyperparameter), is analogous to a speaker ``buffer.''
Then, the speaker-prefixed input for chunk $m$ is given as
\begin{equation}
    \Tilde{\mathbf{X}}_m = \left[\mathbf{B}_1^{\top};\ldots;\mathbf{B}_{K_m}^{\top};\mathbf{X}^{\top}\right]^{\top},
\end{equation}
where $\cdot^{\top}$ denotes transpose.
We use $\Tilde{\mathbf{X}}_m$ instead of $\mathbf{X}_m$ as input for this chunk with the conjecture that the speaker buffers would enforce a relative ordering among speakers in the current chunk.
At the output of the main and auxiliary encoders, we remove the representation corresponding to the prefix, which is of length $\frac{K_m \times \tau}{s}$, where $s$ is the subsampling factor of the encoder.
During inference, we set $\mathscr{S}$ to select a sequence of $\tau$ frames (from the previous chunks) with the largest sum of confidence value, as predicted by its logit $\mathbf{z}^{\mathrm{aux}}[k]$.
%
%
During training, we randomly select $\kappa$ speakers to prefix from all speakers in the batch.
Such a strategy mimics the expected inference time scenario, where not all prefixed speakers will be seen in every chunk.
For each selected speaker, we randomly sample a range of $\tau$ frames from all the segments of that speaker.


\section{Experimental Setup}

\subsection{Network architecture}

The main SURT model follows earlier work~\cite{Raj2023Surt20}.
The masking network comprises four 256-dim DP-LSTM layers~\cite{Luo2019DualPathRE}.
Masked features are reduced to half the original length through a convolutional layer, and the subsampled features are fed into a zipformer encoder~\cite{Yao2023ZipformerAF}.
The ASR encoder consists of 6 zipformer blocks subsampled at different frame rates (up to 8x in the middle).
%
%
The encoder output is further down-sampled such that the overall subsampling factor is 4x.
The representations from an intermediate layer of the ASR encoder are passed to the auxiliary encoder.
This is another zipformer comprising 3 blocks with smaller attention and feed-forward dimensions.
Branch tying is used at the output of both encoders using unidirectional LSTM layers~\cite{Raj2023Surt20}.
The ASR prediction network contains a single 512-dim Conv1D layer.
The complete SURT model contains 38.0M parameters, divided up into 6.0M, 23.6M, and 8.4M for the masking network, the ASR branch, and the speaker branch, respectively. 
The chunk size for the intra-LSTM and the Zipformer is set to 32 frames, resulting in a modeling latency of 320 ms.

\subsection{Data}
\label{sec:data}

\begin{figure}[tb]
    \begin{subfigure}{0.49\linewidth}
    \centering
    \includegraphics[width=\linewidth]{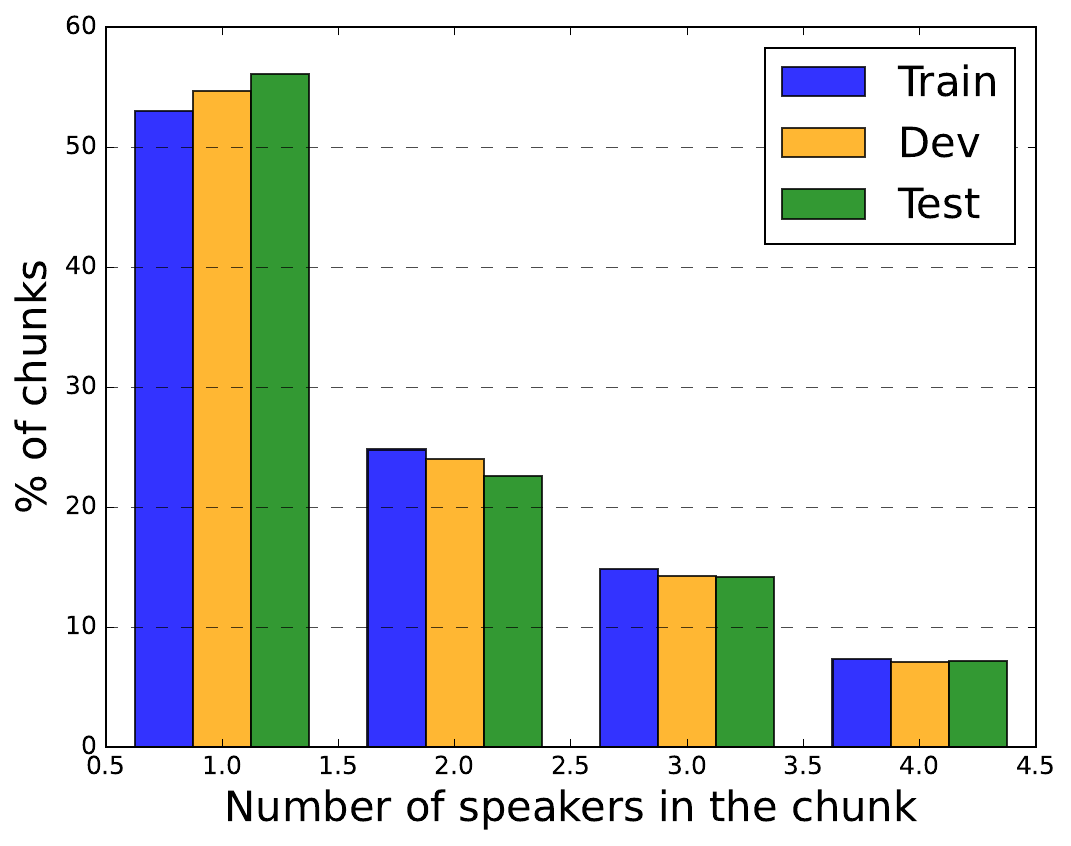}
    \label{fig:spk_chunk}
    \end{subfigure}
    \begin{subfigure}{0.49\linewidth}
    \centering
    \includegraphics[width=\linewidth]{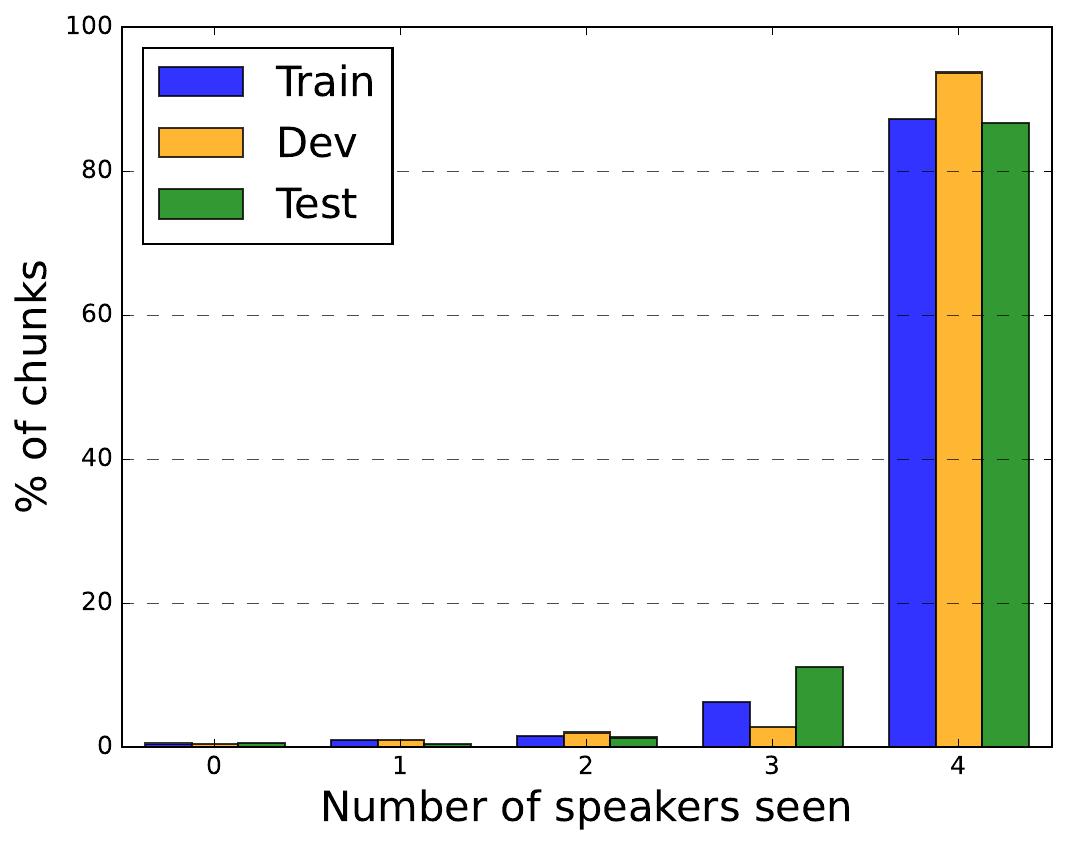}
    \label{fig:spk_seen}
    \end{subfigure}
\vspace{-2em}
\caption{Utterance group statistics of the AMI meeting corpus: (a) number of speakers in the group, and (b) number of speakers seen before the group.}
\label{fig:ami_icsi_stats}
\end{figure}

We conducted our experiments on synthetically mixed LibriSpeech utterances (called LSMix) and the AMI meeting corpus, and their statistics are shown in Table~\ref{tab:stats}.
To create LSMix, we first cut LibriSpeech utterances at 0.2 second pauses, and then mixed speed-perturbed versions of these segments using the algorithm described in \cite{Raj2023Surt20}.
The resulting mixtures were 17s long on average, and contain 2--3 speakers and up to 9 turns of conversation.
We created \texttt{train} and \texttt{dev} splits of LSMix using the corresponding LibriSpeech partitions.
We used this evaluation data to perform ablations for developing the auxiliary speaker branch in SURT.
AMI consists of 100 hours of recorded meetings containing 4 or 5 speakers per session~\cite{Carletta2005TheAM}.
Sessions were recorded on close-talk (headset and lapel) microphones, as well as 2 linear arrays each containing 8 mics.
We used three different mic settings for our experiments: IHM-Mix (digitally mixed individual headset mics), SDM (first channel of array-1), and MDM (beamformed array-1), where the last setting uses officially provided beamformed recordings~\cite{Mir2007AcousticBF}.
To train SURT models for AMI, we used synthetic mixtures of AMI and ICSI~\cite{Janin2003TheIM} utterances (known as AIMix) as described in \cite{Raj2023Surt20}.
We first trained the models on 1841h of the AIMix data, and then adapted them on real \texttt{train} sessions.
For training with the speaker buffer, we fixed $\tau$ as 128 frames for each speaker, and chose $K_m$ from [0,4] with probabilities [0.05,0.05,0.1,0.2,0.6].
This is because during inference, most chunks will be processed with 4 speaker buffers, as seen in Fig.~\ref{fig:ami_icsi_stats}.

\begin{table}[t]
\centering
\caption{Statistics of datasets used for evaluations. The overlap durations are in terms of fraction of total speaking time.}
\label{tab:stats}
\vspace{-1em}
\adjustbox{max width=\linewidth}{
\begin{tabular}{@{}lrrrrr@{}}
\toprule
\multirow{2}{*}{} & \multicolumn{2}{c}{\textbf{LSMix}} & \multicolumn{3}{c}{\textbf{AMI}} \\
\cmidrule(r{2pt}){2-3} \cmidrule(l{2pt}){4-6}
 & \textbf{Train} & \textbf{Dev} & \textbf{Train} & \textbf{Dev} & \textbf{Test} \\ 
\midrule
\textbf{Duration (h:m)} & 2193:57 & 4:19 & 79:23 & 9:40 & 9:03 \\
\textbf{Num. sessions} & 486440 & 897 & 133 & 18 & 16 \\
\textbf{Silence (\%)} & 3.4 & 3.2 & 18.1 & 21.5 & 19.6 \\
\textbf{Overlap (\%)} & 28.4 & 26.0 & 24.5 & 25.7 & 27.0 \\
\bottomrule
\end{tabular}}
\end{table}

\subsection{Training details}
\label{sec:details}

The auxiliary loss scales in \eqref{eq:heat}, $\lambda_{\text{ctc}}$ and $\lambda_{\text{mask}}$, were set to 0.2 each.
We trained the models with the ScaledAdam optimizer following the standard zipformer-transducer recipes in \texttt{icefall}~\cite{Yao2023ZipformerAF}.
This is a variant of Adam where each parameter's update is scaled proportional to the norm of that parameter.
The learning rate was warmed up to 0.004 for 5000 iterations, and decayed exponentially thereafter.
As described in Section~\ref{sec:seq_vs_joint}, we tried sequential and joint training strategies.
For the former, the SURT model was trained for 40 epochs.
For the latter, the ASR branch was first trained for 30 epochs; it was then frozen and the speaker branch was trained for 20 epochs.
In all cases, the ASR transducer was initialized from a pre-trained transducer model, trained for 10 epochs on LibriSpeech, since this has been found useful for fast convergence~\cite{Raj2023Surt20}.
We averaged model checkpoints from the last 5 epochs for inference, and used greedy decoding for reporting all results.
For evaluation on AMI, we initialized the masking network and the ASR branch using the parameters from the SURT model trained on LSMix.
We then trained this model in a sequential process, i.e., ASR branch followed by speaker branch, on AIMix followed by adaptation on real AMI training sessions.

\subsection{Evaluation}
\label{sec:evaluation}

For speaker-agnostic transcription, SURT was evaluated using the optimal reference combination word error rate (ORC-WER) metric, proposed independently in \cite{Sklyar2021MultiTurnRF} and \cite{Raj2021ContinuousSM}.
ORC-WER computes the minimum total WER obtained using the optimal assignment of reference utterances to the output channels.
In this paper, since we have extended SURT to perform speaker-attributed transcription, we measure its performance using the concatenated minimum-permutation WER (cpWER)~\cite{Watanabe2020CHiME6CT}.
This metric finds the best permutation of reference and hypothesis speakers which minimizes the total WER across all speakers.

We also want to measure speaker attribution errors independently of transcription errors. 
The conventional metric for this is known as diarization error rate (DER), and measures the duration ratio of speaking time for which the predicted speakers do not match the reference speakers.
However, since SURT is a streaming model, the ASR tokens and the respective speaker labels may be emitted with some latency compared to their actual reference time-stamp.
This can artificially escalate the DER even when there are few speaker attribution errors.
To circumvent this issue, we report a word-level diarization error rate (WDER) inspired by \cite{Shafey2019JointSR}.
Originally, WDER was defined as the fraction of correctly recognized words which have incorrect speaker tags.
We modify the metric for SURT by using the ORC-WER reference assignment to identify the correct words and the speaker mapping from the cpWER computation to check for speaker equivalence.


\section{Results \& Discussion}



\subsection{RNN-T vs. HAT for speaker-agnostic ASR}

Since our formulation requires replacing the conventional RNN-T loss, i.e. \eqref{eq:rnnt_softmax}, with the HAT loss given by \eqref{eq:hat}, we want to ensure that the speaker-agnostic ASR performance of the model does not degrade.
To verify this, we trained SURT (without an auxiliary branch) using $\mathcal{L}_{\mathrm{rnnt}}$ and $\mathcal{L}_{\mathrm{hat}}$ on the LSMix \texttt{train} set, and evaluated the resulting models on the \texttt{dev} set.
We found that the \textbf{HAT model obtained 8.53\% ORC-WER}, compared to 8.59\% using regular RNN-T.
The error breakdown showed marginally higher insertions but fewer deletions, which may be due to explicit modeling of the blank token.

\begin{table}[t]
\centering
\caption{Comparison of different training strategies for SURT with auxiliary speaker branch.}
\label{tab:training_strategy}
\vspace{-1em}
\adjustbox{max width=\linewidth}{
\begin{tabular}{@{}lrrr@{}}
\toprule
\textbf{Strategy} & \textbf{ORC-WER} & \textbf{WDER} & \textbf{cpWER} \\ \midrule
Sequential & 8.53 & \textbf{3.99} & 14.96 \\
Joint & \textbf{8.43} & 4.46 & \textbf{14.95} \\
Seq. + Joint & 9.17 & 4.25 & 15.33 \\
\bottomrule
\end{tabular}
}
\vspace{0.5em}
\end{table}

\subsection{Sequential vs. joint training}
\label{sec:seq_vs_joint}

The auxiliary speaker branch of the SURT model can be trained in several ways, as shown in Table~\ref{tab:training_strategy}.
In ``sequential'' training, the main SURT model is first trained using \eqref{eq:heat} and then frozen while the auxiliary branch is trained using \eqref{eq:spk_hat}.
In ``joint'' training, the full model is trained from scratch with the multi-task objective.
Finally, we can combine the above approaches by first training the branches sequentially and then fine-tuning them jointly.
We found that \textbf{both sequential and joint training resulted in similar cpWER performance}, but joint training degrades WDER.
Furthermore, joint fine-tuning after sequential training degraded performance on both ASR metrics.
Since sequential training allows decoupling of ASR and speaker attribution performance, we used this strategy for the experiments in the remainder of this paper.

\subsection{Auxiliary encoder position}

\begin{table}[t]
\centering
\caption{Speaker-attributed ASR performance on LSMix \texttt{dev} for different positions of the auxiliary encoder. $\mathbf{h}_l$ denotes the hidden representation at the $l^{\mathrm{th}}$ block of the main zipformer encoder, and $\mathbf{h}_0^{\mathrm{aux}}$ is the input to the auxiliary encoder.}
\label{tab:aux_position}
\vspace{-1em}
\adjustbox{max width=\linewidth}{
\begin{tabular}{@{}lcccrr@{}}
\toprule
$\mathbf{h}_{0}^{\mathrm{aux}}$ & \textbf{Ins.} & \textbf{Del.} & \textbf{Sub.} & \textbf{cpWER} & \textbf{WDER} \\ \midrule
$=\mathbf{h}_{0}$ & 3.34 & 6.04 & 7.28 & 16.66 & 5.36 \\
$=\mathbf{h}_{1}$ & \textbf{2.91} & \textbf{5.20} & \textbf{6.85} & \textbf{14.96} & \textbf{3.99} \\
$=\mathbf{h}_{2}$ & 4.58 & 6.77 & 8.24 & 19.59 & 6.73 \\
$=\mathbf{h}_{3}$ & 5.95 & 8.23 & 9.41 & 23.59 & 8.35 \\
\bottomrule
\end{tabular}
}
\end{table}

The input $\mathbf{h}_{0}^{\mathrm{aux}}$ to the auxiliary encoder is obtained from an intermediate representation of the main encoder.
We trained several SURT models with different positions for the auxiliary input, in order to find the optimal representation for the speaker branch, and the results are shown in Table~\ref{tab:aux_position}.
The models were trained sequentially and the ORC-WER was 8.53\% (same as earlier).
We found that both cpWER and WDER get progressively worse if we used representations from deeper layers, possibly because of loss in speaker information through the main encoder.
Interestingly, $\mathbf{h}_1$ (i.e. output of the first zipformer block) showed better performance than $\mathbf{h}_0$ (output from convolutional embedding layer).
We conjecture that the input to the \textbf{auxiliary encoder needs contextualized representations} since speaker labels need to be synchronized across the two branches.
These findings mirror recent studies showing that intermediate layers of the acoustic model are most suitable for extracting speaker information~\cite{Huang2023TowardsWE}.
Such analysis has also motivated ``tandem'' multi-task learning of ASR and speaker diarization using self-supervised encoders such as Wav2Vec 2.0~\cite{Zheng2022TandemMT}.

\begin{figure}[tb]
    \begin{subfigure}{0.49\linewidth}
    \centering
    \includegraphics[width=\linewidth]{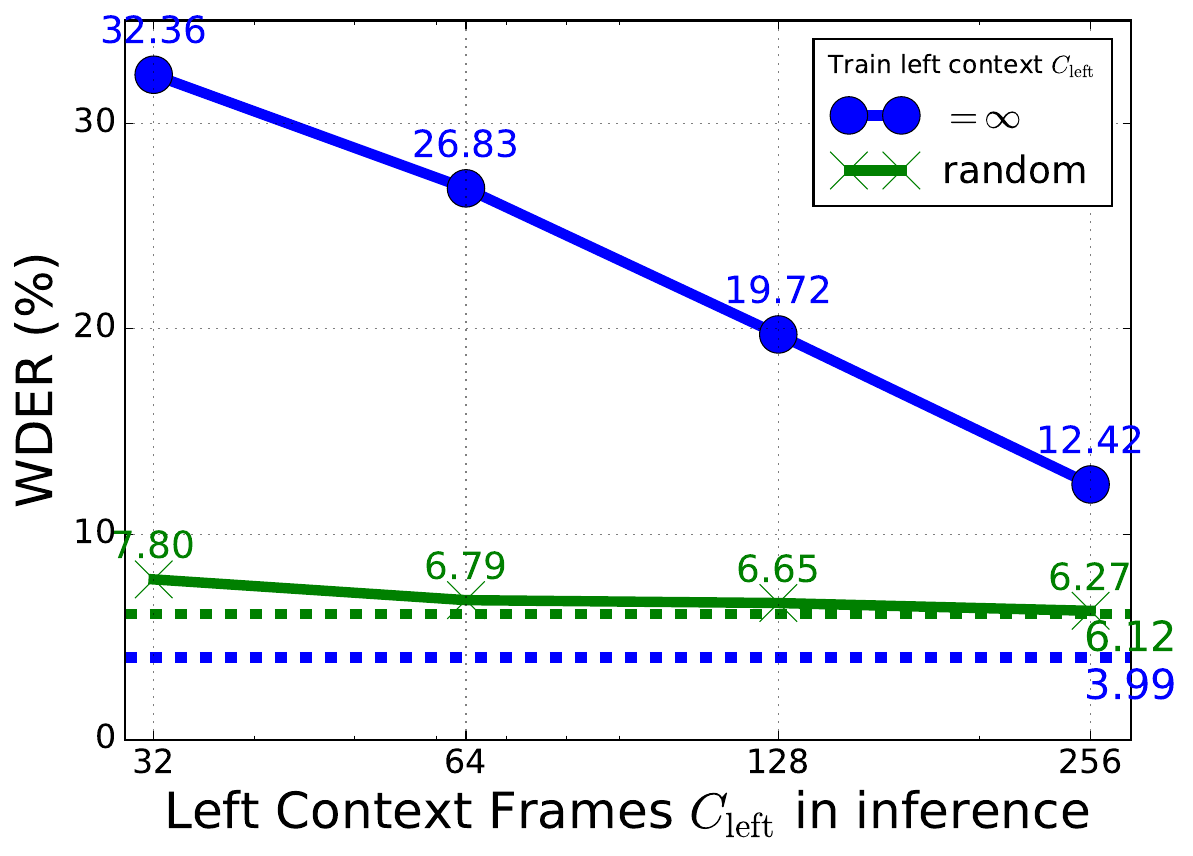}
    \label{fig:lc_der}
    \end{subfigure}
    \begin{subfigure}{0.49\linewidth}
    \centering
    \includegraphics[width=\linewidth]{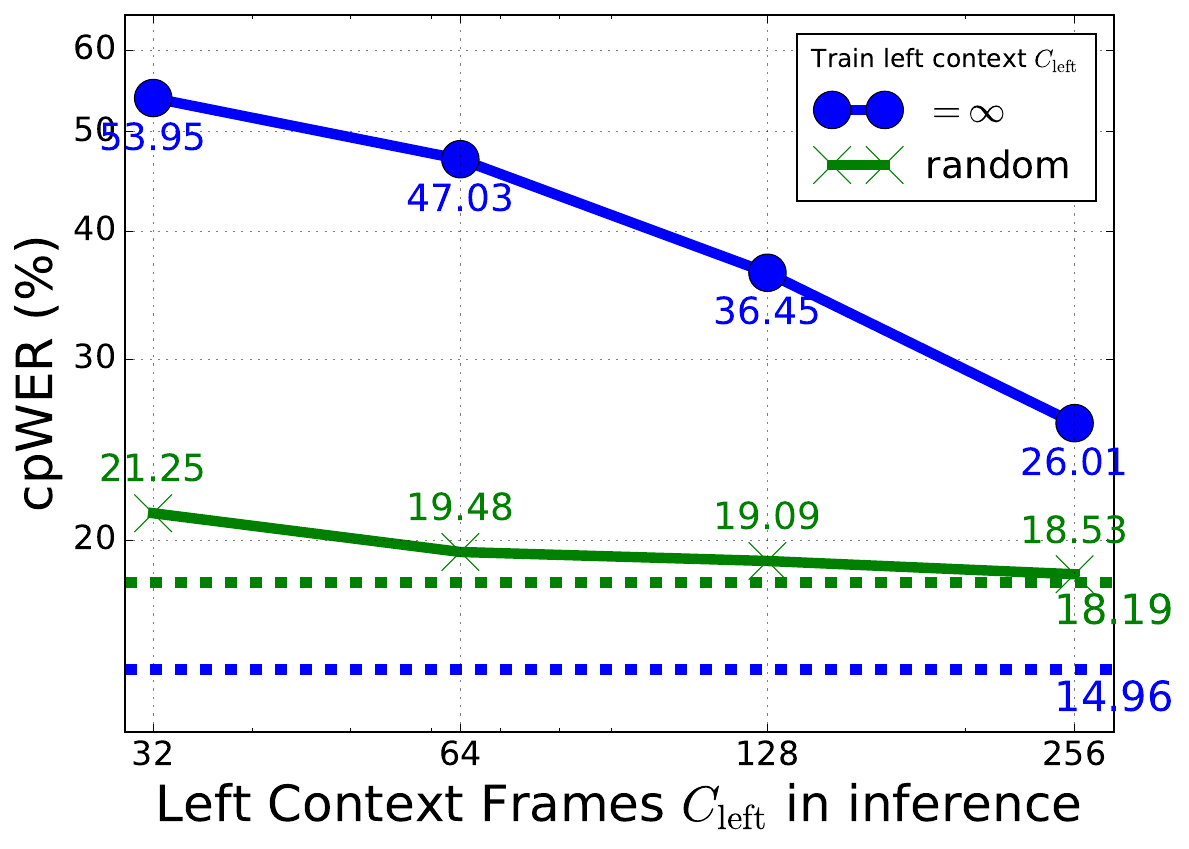}
    \label{fig:lc_wer}
    \end{subfigure}
\vspace{-2em}
\caption{Effect of auxiliary encoder left context on (a) WDER and (b) cpWER. Dotted lines show best performance using $\infty$ left context.}
\label{fig:left_context}
\end{figure}

\subsection{Effect of left context}

The ASR encoder of the SURT model uses limited left context ($C_{\mathrm{left}}$=128 frames) in the self-attention computation during inference.
While ASR token prediction is usually a local decision, speaker label prediction requires looking at the full history in order to synchronize the relative FIFO labels.
We experimented with training and decoding with different histories, and the results are shown in Fig.~\ref{fig:left_context}.
For a model trained with infinite $C_{\mathrm{left}}$ (solid blue line), limiting it during inference quickly degraded WDER and cpWER performance.
When the model was trained with randomized $C_{\mathrm{left}}$ (solid green line), the degradation was less evident.
However, it was unable to make full use of infinite history at inference time, and only obtained a WDER of 6.12\%, versus 3.99\% for the model trained with infinite $C_{\mathrm{left}}$.
This indicates that using \textbf{infinite left context during training and inference is important} for the auxiliary speaker encoder.

\begin{table*}[t]
\begin{minipage}{.65\linewidth}
\centering
\caption{Performance of SURT models (with and without speaker prefixing) for different conditions on AMI \texttt{test} set, evaluated on utterance groups. ``ORC'' denotes the ORC-WER metric, and is the same for all models.}
\label{tab:ami_result}
\vspace{-1em}
\adjustbox{max width=\linewidth}{
\begin{tabular}{@{}lcccccccccc@{}}
\toprule
\multicolumn{2}{c}{\textbf{Prefix}} &
\multicolumn{3}{c}{\textbf{IHM-Mix}} & \multicolumn{3}{c}{\textbf{SDM}} & \multicolumn{3}{c}{\textbf{MDM}} \\
\cmidrule(r{5pt}){1-2} \cmidrule(l{2pt}r{2pt}){3-5} \cmidrule(l{2pt}r{2pt}){6-8} \cmidrule(l{4pt}){9-11}
\textbf{ID} & \textbf{Train/Decode} & ORC & WDER & cpWER & ORC & WDER & cpWER & ORC & WDER & cpWER \\
\midrule
\textbf{A} & \xmark ~/~  \xmark & 34.9 & \textbf{9.3} & \textbf{42.9} & 43.2 & \textbf{10.9} & \textbf{50.3} & 40.5 & \textbf{9.9} & \textbf{47.3} \\
\textbf{B} & \xmark ~/~ \cmark & 34.9 & 22.5 & 61.2 & 43.2 & 23.1 & 68.2 & 40.5 & 22.6 & 64.8 \\
\textbf{C} & \cmark ~/~ \cmark & 34.9 & 14.0 & 49.9 & 43.2 & 16.3 & 58.9 & 40.5 & 15.5 & 56.0 \\
\bottomrule
\end{tabular}}
\end{minipage}%
\hfill
\begin{minipage}{.32\linewidth}
\centering
\caption{Breakdown of model (A)'s performance on IHM-Mix \texttt{test} set by number of speakers in the utterance group.}
\label{tab:ami_num_spk}
\vspace{-1em}
\adjustbox{max width=\linewidth}{
\begin{tabular}{@{}lrrrrr@{}}
\toprule
\textbf{\#spk} & \textbf{1} & \textbf{2} & \textbf{3} & \textbf{4} & \textbf{Avg.} \\ \midrule
WDER ($\downarrow$) & 0.1 & 3.4 & 13.0 & 23.9 & 9.3 \\
Count. ($\uparrow$) & 98.6 & 61.9 & 26.8 & 44.0 & 75.9 \\
cpWER ($\downarrow$) & 17.2 & 32.4 & 51.1 & 63.6 & 42.9 \\
\bottomrule
\end{tabular}}
\end{minipage}
\vspace{-1.5em}
\end{table*}

\subsection{Utterance-group evaluation on AMI}

We evaluated the SURT model on different microphone settings of the AMI meeting corpus in the utterance-group scenario, and the results are shown in Table~\ref{tab:ami_result} in terms of ORC-WER, WDER, and cpWER.
%
%
Across the board, performance \textbf{degraded from IHM-Mix to SDM} settings, which is expected since SDM contains far-field artifacts in addition to overlaps.
Beamforming with multiple microphones partially removes background noise and reverberations, thus providing a slightly easier condition than SDM.
For system A, which was trained and decoded without speaker prefixing, we obtained a cpWER of 46.8\%, on average across the three conditions.
When we used the same model for decoding with speaker prefixes (system B), the cpWER performance degraded by 38.2\% relative to the former.
Since the model has not seen short speaker buffers at train time, the auxiliary encoder is not adept at using these for generating the contextualized representations.

\begin{figure}[t]
    \centering
    \includegraphics[width=\linewidth]{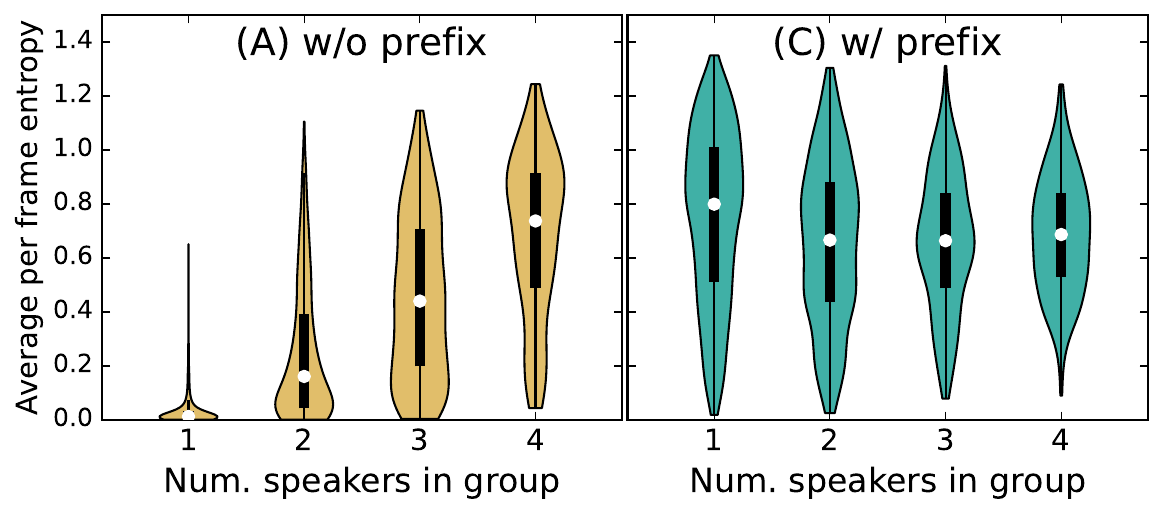}
    \vspace{-2em}
    \caption{Average per-frame entropy for utterance groups with different number of speakers.}
    \label{fig:entropy}
\end{figure}

Next, we trained the same SURT model using speaker prefixing as described in Section~\ref{sec:details}, and found that it improved performance significantly due to matched train and test conditions.
Nevertheless, this model was \textbf{7-8\% worse than the original} model in terms of absolute cpWER performance.
To investigate this further, we computed the average framewise entropy over speaker labels for all utterance groups in the IHM-Mix \texttt{test} set, and grouped them by number of speakers in the group.
Fig.~\ref{fig:entropy} shows the distribution of these entropies for the SURT model with and without speaker prefixing.
We found that for the model without prefixing, the \textbf{entropy was very low for utterance groups with a single speaker}, and gradually increased with the number of speakers.
This indicates that the model was very confident in its prediction for few speaker cases.
The opposite trend was seen for the model with speaker prefixing, where the entropy was highest for the single-speaker case.
This is because for each frame, the model needs to decide which of the 4 prefixed speakers the frame should be assigned to, which may result in low confidence of prediction.
%

In general, we found that the performance of all models  \textbf{gets progressively worse} as the number of speakers in the group increases, as shown in Table~\ref{tab:ami_num_spk} for system A.
Interestingly, the degradation in WDER was small compared to that in the speaker counting accuracy.
This may be because of several utterance groups where some speakers participate with just a few words, which may be hard for the system to identify, but do not contribute much in overall speaker attribution error.

\subsection{Full-session evaluation on AMI}

Finally, we performed inference on full AMI \texttt{test} sessions and the corresponding cpWERs are reported n Table~\ref{tab:ami_result2}.
Computing the ORC-WER and WDER for this case was not feasible since their computational complexity depends on the number of segments in the reference.
First, we see that the model without speaker prefixing obtained very high error rates, since it failed at correctly reconciling speaker labels across different utterance groups.
With speaker prefixing, we obtained \textbf{15.1\% relative cpWER improvement} on average across the mic settings.
For the speaker prefixing, we trained and evaluated the model using $\tau$ of 128 frames or 1.28s per speaker.

In a meeting transcription setup, since the participants are known before-hand, we can usually obtain an enrollment utterance for each speaker.
Instead of selecting speaker prefixes from previous chunks, if we select them from these enrollment utterances, we obtain a further \textbf{relative cpWER improvement of 29.2\%}, on average.
We conjecture that when enrollment utterances are not used, speaker attribution errors in earlier chunks can adversely impact performance on current chunk, since the buffer frames are used to guide the relative order.
Nevertheless, there still exists a significant gap of about 10--12\% absolute cpWER between full session evaluation and utterance group evaluation (shown in Table~\ref{tab:ami_result}).

\begin{table}[t]
\centering
\caption{Full-session evaluation cpWER (\%) on AMI \texttt{test} set.}
\label{tab:ami_result2}
\vspace{-1em}
\adjustbox{max width=\linewidth}{
\begin{tabular}{@{}lrrr@{}}
\toprule
\textbf{Method} & \textbf{IHM-Mix} & \textbf{SDM} & \textbf{MDM} \\
\midrule
w/o prefix & 100.11 & 97.15 & 96.26 \\
w/ prefix & 82.77 & 83.94 & 82.28 \\
~~+ enrollment & 53.04 & 61.22 & 59.59  \\
\bottomrule
\end{tabular}}
\end{table}

\section{Conclusion}

The SURT framework allows continuous, streaming recognition of multi-talker conversations, but it could only be used for speaker-agnostic transcription.
In this paper, we showed how to perform streaming word-level speaker labeling with SURT, thus enabling speaker-attributed transcription using the same model.
We achieved this by adding an auxiliary speaker encoder to the recognition component of the model, and used the same two-branch strategy to handle overlapping speech.
We solved the problem of synchronization between the ASR and speaker branch outputs by factoring out the blank logit and sharing it between the branches.
Since the model predicts relative speaker labels in FIFO order, reconciling the labels across utterance groups in a recording becomes a challenge.
We showed that a simple strategy of prefixing high-confidence speaker frames for the recognized speakers can partially alleviate this problem, but it would require further investigation to bring session-level error rates closer to those for utterance groups.

\bibliographystyle{IEEEbib_abbrev}
\bibliography{main}

\begin{thebibliography}{10}

\bibitem{Barker2015TheT}
J. Barker, R. Marxer, E. Vincent, and S. Watanabe,
\newblock ``The third {‘CHiME’} speech separation and recognition challenge: Dataset, task and baselines,''
\newblock in {\em IEEE ASRU}, 2015.

\bibitem{Kinoshita2013TheRC}
K. Kinoshita, M. Delcroix, T. Yoshioka, T. Nakatani, A. Sehr, W. Kellermann, and R. Maas,
\newblock ``The {REVERB} challenge: A common evaluation framework for dereverberation and recognition of reverberant speech,''
\newblock in {\em IEEE WASPAA}, 2013.

\bibitem{Watanabe2020CHiME6CT}
S. Watanabe, M. Mandel, J. Barker, and E. Vincent,
\newblock ``{CHiME-6} challenge: Tackling multispeaker speech recognition for unsegmented recordings,''
\newblock {\em ArXiv}, 2020.

\bibitem{Carletta2005TheAM}
J. Carletta et~al.,
\newblock ``The {AMI} meeting corpus: A pre-announcement,''
\newblock in {\em MLMI}, 2005.

\bibitem{Shriberg2001ObservationsOO}
E. Shriberg, A. Stolcke, and D. Baron,
\newblock ``Observations on overlap: findings and implications for automatic processing of multi-party conversation,''
\newblock in {\em Interspeech}, 2001.

\bibitem{Yoshioka2019MeetingTU}
T. Yoshioka, D. Dimitriadis, A. Stolcke, W. Hinthorn, Z. Chen, M. Zeng, and X. Huang,
\newblock ``Meeting transcription using asynchronous distant microphones,''
\newblock in {\em Interspeech}, 2019.

\bibitem{Raj2022GPUacceleratedGS}
D. Raj, D. Povey, and S. Khudanpur,
\newblock ``{GPU}-accelerated guided source separation for meeting transcription,''
\newblock in {\em Interspeech}, 2023.

\bibitem{Raj2020IntegrationOS}
D. Raj, P. Denisov, et~al.,
\newblock ``Integration of speech separation, diarization, and recognition for multi-speaker meetings: System description, comparison, and analysis,''
\newblock in {\em IEEE SLT}, 2021.

\bibitem{Kanda2019SimultaneousSR}
N. Kanda, S. Horiguchi, Y. Fujita, Y. Xue, K. Nagamatsu, and S. Watanabe,
\newblock ``Simultaneous speech recognition and speaker diarization for monaural dialogue recordings with target-speaker acoustic models,''
\newblock in {\em IEEE ASRU}, 2019, pp. 31--38.

\bibitem{Wu2021InvestigationOP}
J. Wu, Z. Chen, S. Chen, Y. Wu, T. Yoshioka, N. Kanda, S. Liu, and J. Li,
\newblock ``Investigation of practical aspects of single channel speech separation for {ASR},''
\newblock in {\em Interspeech}, 2021.

\bibitem{Chorowski2015AttentionBasedMF}
J. Chorowski, D. Bahdanau, D. Serdyuk, K. Cho, and Y. Bengio,
\newblock ``Attention-based models for speech recognition,''
\newblock in {\em NIPS}, 2015.

\bibitem{Kanda2020SerializedOT}
N. Kanda, Y. Gaur, X. Wang, Z. Meng, and T. Yoshioka,
\newblock ``Serialized output training for end-to-end overlapped speech recognition,''
\newblock in {\em Interspeech}, 2020.

\bibitem{Kanda2020JointSC}
N. Kanda, Y. Gaur, et~al.,
\newblock ``Joint speaker counting, speech recognition, and speaker identification for overlapped speech of any number of speakers,''
\newblock in {\em Interspeech}, 2020.

\bibitem{Kanda2021EndtoEndSA}
N. Kanda, G. Ye, Y. Gaur, X. Wang, Z. Meng, Z. Chen, and T. Yoshioka,
\newblock ``End-to-end speaker-attributed {ASR} with transformer,''
\newblock in {\em Interspeech}, 2021.

\bibitem{Kanda2021ACS}
N. Kanda, X. Xiao, J. Wu, T. Zhou, Y. Gaur, X. Wang, Z. Meng, Z. Chen, and T. Yoshioka,
\newblock ``A comparative study of modular and joint approaches for speaker-attributed asr on monaural long-form audio,''
\newblock in {\em IEEE ASRU}, 2021, pp. 296--303.

\bibitem{Chang2021HypothesisSF}
X. Chang, N. Kanda, Y. Gaur, X. Wang, Z. Meng, and T. Yoshioka,
\newblock ``Hypothesis stitcher for end-to-end speaker-attributed {ASR} on long-form multi-talker recordings,''
\newblock in {\em IEEE ICASSP}, 2021.

\bibitem{Kanda2020InvestigationOE}
N. Kanda, X. Chang, Y. Gaur, X. Wang, Z. Meng, Z. Chen, and T. Yoshioka,
\newblock ``Investigation of end-to-end speaker-attributed {ASR} for continuous multi-talker recordings,''
\newblock in {\em IEEE SLT}, 2020.

\bibitem{Yu2022ACS}
F. Yu, Z. Du, S. Zhang, Y. Lin, and L. Xie,
\newblock ``A comparative study on speaker-attributed automatic speech recognition in multi-party meetings,''
\newblock in {\em Interspeech}, 2022.

\bibitem{Shi2022ACS}
M. Shi, J. Zhang, Z. Du, F. Yu, S. Zhang, and L. Dai,
\newblock ``A comparative study on multichannel speaker-attributed automatic speech recognition in multi-party meetings,''
\newblock in {\em APSIPA ASC}, 2022, pp. 1943--1948.

\bibitem{Shi2023CASAASRCS}
M. Shi, Z. Du, Q. Chen, F. Yu, Y. Li, S. Zhang, J. Zhang, and L. Dai,
\newblock ``Casa-asr: Context-aware speaker-attributed asr,''
\newblock in {\em Interspeech}, 2023.

\bibitem{Kanda2022StreamingSA}
N. Kanda, J. Wu, Y. Wu, X. Xiao, Z. Meng, X. Wang, Y. Gaur, Z. Chen, J. Li, and T. Yoshioka,
\newblock ``Streaming speaker-attributed {ASR} with token-level speaker embeddings,''
\newblock in {\em Interspeech}, 2022.

\bibitem{Graves2012SequenceTW}
A. Graves,
\newblock ``Sequence transduction with recurrent neural networks,''
\newblock in {\em ICML Representation Learning Workshop}, 2012.

\bibitem{Raj2023Surt20}
D. Raj, D. Povey, and S. Khudanpur,
\newblock ``{SURT} 2.0: Advances in transducer-based multi-talker speech recognition,''
\newblock {\em IEEE/ACM Transactions on Audio, Speech, and Language Processing}, vol. 31, pp. 3800--3813, 2023.

\bibitem{Lu2020StreamingEM}
L. Lu, N. Kanda, J. Li, and Y. Gong,
\newblock ``Streaming end-to-end multi-talker speech recognition,''
\newblock {\em IEEE Signal Processing Letters}, vol. 28, pp. 803--807, 2020.

\bibitem{Sklyar2021MultiTurnRF}
I. Sklyar, A. Piunova, X. Zheng, and Y. Liu,
\newblock ``Multi-turn {RNN-T} for streaming recognition of multi-party speech,''
\newblock in {\em IEEE ICASSP}, 2021.

\bibitem{Raj2021ContinuousSM}
D. Raj, L. Lu, Z. Chen, Y. Gaur, and J. Li,
\newblock ``Continuous streaming multi-talker {ASR} with dual-path transducers,''
\newblock in {\em IEEE ICASSP}, 2022.

\bibitem{Lu2022EndpointDF}
L. Lu, J. Li, and Y. Gong,
\newblock ``Endpoint detection for streaming end-to-end multi-talker {ASR},''
\newblock in {\em IEEE ICASSP}, 2022.

\bibitem{Sklyar2022SeparatorTransducerSegmenterSR}
I. Sklyar, A. Piunova, and C. Osendorfer,
\newblock ``Separator-transducer-segmenter: Streaming recognition and segmentation of multi-party speech,''
\newblock in {\em Interspeech}, 2022.

\bibitem{Lu2021StreamingMS}
L. Lu, N. Kanda, J. Li, and Y. Gong,
\newblock ``Streaming multi-talker speech recognition with joint speaker identification,''
\newblock in {\em Interspeech}, 2021.

\bibitem{Kuang2022PrunedRF}
F. Kuang, L. Guo, W. Kang, L. Lin, M. Luo, Z. Yao, and D. Povey,
\newblock ``Pruned {RNN-T} for fast, memory-efficient {ASR} training,''
\newblock in {\em Interspeech}, 2022.

\bibitem{Graves2006ConnectionistTC}
A. Graves, S. Fern{\'a}ndez, F. Gomez, and J. Schmidhuber,
\newblock ``Connectionist temporal classification: labelling unsegmented sequence data with recurrent neural networks,''
\newblock in {\em ICML}, 2006.

\bibitem{Flemotomos2019LanguageAS}
N. Flemotomos, P.~G. Georgiou, and S.~S. Narayanan,
\newblock ``Language aided speaker diarization using speaker role information,''
\newblock in {\em Speaker Odyssey}, 2020.

\bibitem{Park2019SpeakerDW}
T.~J. Park, K.~J. Han, J. Huang, X. He, B. Zhou, P.~G. Georgiou, and S.~S. Narayanan,
\newblock ``Speaker diarization with lexical information,''
\newblock in {\em Interspeech}, 2019.

\bibitem{Khare2022ASRAwareEN}
A. Khare, E. Han, Y. Yang, and A. Stolcke,
\newblock ``Asr-aware end-to-end neural diarization,''
\newblock in {\em IEEE ICASSP}, 2022, pp. 8092--8096.

\bibitem{Variani2020HybridAT}
E. Variani, D. Rybach, C. Allauzen, and M. Riley,
\newblock ``Hybrid autoregressive transducer (hat),''
\newblock in {\em IEEE ICASSP}, 2020, pp. 6139--6143.

\bibitem{Huang2023TowardsWE}
Y. Huang, W. Wang, G. Zhao, H. Liao, W. Xia, and Q. Wang,
\newblock ``Towards word-level end-to-end neural speaker diarization with auxiliary network,''
\newblock {\em ArXiv}, vol. abs/2309.08489, 2023.

\bibitem{Kinoshita2020IntegratingEN}
K. Kinoshita, M. Delcroix, and N. Tawara,
\newblock ``Integrating end-to-end neural and clustering-based diarization: Getting the best of both worlds,''
\newblock in {\em IEEE ICASSP}, 2020, pp. 7198--7202.

\bibitem{Xue2020OnlineEN}
Y. Xue, S. Horiguchi, Y. Fujita, S. Watanabe, and K. Nagamatsu,
\newblock ``Online end-to-end neural diarization with speaker-tracing buffer,''
\newblock in {\em IEEE SLT}, 2021, pp. 841--848.

\bibitem{Luo2019DualPathRE}
Y. Luo, Z. Chen, and T. Yoshioka,
\newblock ``Dual-path {RNN}: Efficient long sequence modeling for time-domain single-channel speech separation,''
\newblock in {\em IEEE ICASSP}, 2019.

\bibitem{Yao2023ZipformerAF}
Z. Yao, L. Guo, X. Yang, W. Kang, F. Kuang, Y. Yang, Z. Jin, L. Lin, and D. Povey,
\newblock ``Zipformer: A faster and better encoder for automatic speech recognition,''
\newblock {\em ArXiv}, vol. abs/2310.11230, 2023.

\bibitem{Mir2007AcousticBF}
X.~A. Mir{\'o}, C. Wooters, and J. Hernando,
\newblock ``Acoustic beamforming for speaker diarization of meetings,''
\newblock {\em IEEE TASLP}, vol. 15, pp. 2011--2022, 2007.

\bibitem{Janin2003TheIM}
A.~L. Janin, D. Baron, J. Edwards, D.~P.~W. Ellis, D. Gelbart, N. Morgan, B. Peskin, T. Pfau, E. Shriberg, A. Stolcke, and C. Wooters,
\newblock ``The {ICSI} meeting corpus,''
\newblock in {\em IEEE ICASSP}, 2003.

\bibitem{Shafey2019JointSR}
L.~E. Shafey, H. Soltau, and I. Shafran,
\newblock ``Joint speech recognition and speaker diarization via sequence transduction,''
\newblock in {\em Interspeech}, 2019.

\bibitem{Zheng2022TandemMT}
X. Zheng, C. Zhang, and P.~C. Woodland,
\newblock ``Tandem multitask training of speaker diarisation and speech recognition for meeting transcription,''
\newblock in {\em Interspeech}, 2022.

\end{thebibliography}

\end{document}